\documentstyle[twocolumn,prl,aps,floats,psfig]{revtex} 
\newcommand{\bsim}{\mbox{\raisebox{-0.1cm}{$\;
\stackrel{\textstyle>}{\sim}\;$}}}
\newcommand{\lsim}{\mbox{\raisebox{-0.1cm}{$\;
\stackrel{\textstyle<}{\sim}\;$}}}

\begin{document}
\twocolumn[\hsize\textwidth\columnwidth\hsize\csname 
@twocolumnfalse\endcsname

\title{Nonadiabatic Channels in the Superconducting Pairing
of Fullerides} 

\author{E. Cappelluti$^1$, C. Grimaldi$^2$, L. Pietronero$^1$, and
S. Str\"assler$^2$} 

\address{$^1$ Dipartimento di Fisica, Universit\'{a} di Roma 
``La Sapienza", 
Piazzale A.  Moro, 2, 00185 Roma, Italy \\
and Istituto Nazionale Fisica della Materia, Unit\'a di Roma 1, Italy}
\address{$^2$ \'Ecole Polytechnique F\'ed\'erale de Lausanne,
D\'epartement de Microtechnique IPM, CH-1015 Lausanne, Switzerland}

\date{\today} 
\maketitle 

\begin{abstract}
We show the intrinsic inconsistency of the conventional
phonon mediated theory of superconductivity in relation
to the observed properties of Rb$_3$C$_{60}$. The recent, highly accurate
measurement of the carbon isotope coefficient $\alpha_{\rm C}=0.21$,
together with the high value of $T_c$ ($30$ K) and the very small
Fermi energy $E_{\rm F}$ ($0.25$ eV), unavoidably
implies the opening of nonadiabatic channels in the superconducting
pairing. We estimate these effects and show that they are actually
the key elements for the high value of $T_c$ in these materials
compared to the very low values of graphite intercalation compounds.
\\
PACS number(s): 74.20.-z, 74.70.Wz, 71.38.+i
\end{abstract}

\vskip 2pc]

\narrowtext

One of the most striking evidences of the phonon role
in high-temperature superconductivity of alkali-doped C$_{60}$
compounds is the observation of nonzero isotope effect on the value
of the critical temperature $T_c$ \cite{gunnarsson}. 
However, the large spread of
the reported values of the carbon isotope coefficient
$\alpha_{\rm C}=-d\ln T_c/d\ln M$ \cite{iso},
where $M$ is the isotopic mass,
has prevented to settle a definitive and self-consistent picture.
This long standing uncertainty has
been solved only recently for the compound Rb$_3$C$_{60}$
($T_c\simeq 30$ K). Resistive measurements on $99$\%
enriched $^{13}$C single-crystals have permitted in fact to
establish $\alpha_{\rm C}=0.21$ with high precision \cite{fuhrer}.

The knowledge of the accurate value of $\alpha_{\rm C}$ is an important
element to establish not only the important role of phonons, but
also to test the self-consistency of the Migdal-Eliashberg (ME)
theory of superconductivity \cite{migdal,elia} in
alkali-doped fullerenes. The measured values $\alpha_{\rm C}=0.21$
and $T_c=30$ K can be used to extract the
microscopic quantities involved in the superconducting pairing.
For example, in Ref. \cite{fuhrer}, 
$T_c=30$ K and $\alpha_{\rm C}=0.21$ are interpreted within the 
conventional ME theory by $\lambda=0.9$, $\omega_{\rm ln}=1360$ K and 
$\mu^*=0.22$, where $\lambda$ is the 
electron-phonon coupling constant, $\omega_{\rm ln}$ is the logarithmic
phonon frequency and $\mu^*$ is the Coulomb pseudopotential \cite{carbotte}.
According to this standard analysis, the high value of $T_c$
in alkali-doped C$_{60}$ compounds is merely due to
a strong electron-phonon
coupling to the highest intramolecular phonon modes.
These results should be compared with the graphite intercalation compounds
(GIC) where $T_c \simeq 0.2$  K \cite{belash} is explained by a moderate coupling 
($\lambda \simeq 0.3$) to similar high energy phonon modes.
Current theories claim that the big difference between
the electron-phonon coupling in fullerides compared with
graphite intercalation compounds arises from the finite curvature
of the C$_{60}$ molecule \cite{benning}. 
In this perspective, therefore, Rb$_3$C$_{60}$ is just
an ordinary strong-coupling superconductor 
described by the conventional adiabatic ME framework.

In this Letter instead we demonstrate the intrinsic inconsistency
of the standard ME theory in Rb$_3$C$_{60}$. This conventional
description is in fact invalidated by the extremely low value
of the Fermi energy $E_{\rm F}\simeq 0.25$ eV $\simeq 2900$ K
 characteristic
of the fullerene compounds \cite{gunnarsson}.
We show that the whole range of 
$\lambda$-$\omega_{\rm ph}$ values which fit $T_c=30$ K and 
$\alpha_{\rm C}=0.21$ through the solution of the standard ME equations 
implies a Migdal parameter $\lambda\omega_{\rm ph}/E_{\rm F}$ larger
than $0.4$, instead of being zero as assumed by the ME theory \cite{migdal,elia}.
This situation inevitably leads to the breakdown of Migdal's
theorem and to the opening of nonadiabatic channels in the superconducting
pairing \cite{gpsprl}. By solving the nonadiabatic equations \cite{psg,sgp}, 
we estimate these effects and show that they are actually the
key elements for the high values of $T_c$ in the fullerene compounds.

We now discuss why the experimental data $T_c=30$ K and
$\alpha_{\rm C}=0.21$ \cite{fuhrer} are inconsistent with respect to the ME
theory.
To simulate the coupling of the electrons to the different intramolecular
phonon modes, we consider an electron-phonon spectral function
$\alpha^2\! F(\omega)$ modeled as a rectangle of width $\Delta\omega_0$
and centered at $\omega_0$. For $\Delta\omega_0=0$, $\alpha^2\! F(\omega)$
reduces to a single Einstein $\delta$-peak while for $\Delta\omega_0>0$
it becomes broadened.
The electron-phonon coupling constant
is determined by the usual relation $\lambda=2\int d\omega
\alpha^2\! F(\omega)/\omega$. 
The Coulomb pseudopotential is taken to have the
standard form $\mu^*=\mu/[1+\mu\ln(\omega_c/\omega_{\rm max})]$, where
$\omega_{\rm max}=\omega_0+\Delta\omega_0/2$ is the maximum phonon
frequency, $\mu$ is the screened Coulomb parameter and $\omega_c$ is the 
high-frequency cut-off.
According to whether it is the entire set of $\pi$-bands \cite{varma} or rather only
the narrow conducting $t_{\rm 1u}$ band \cite{koch} which contributes to the
dynamical screening, we shall consider $\omega_c=5\omega_{\rm max}$
or $\omega_c=E_{\rm F}=2900$K, respectively.
For different values of $\Delta\omega_0/\omega_0$ we then solve
numerically the
ME equations to find the values of $\lambda$, $\omega_0$, and $\mu$ (or $\mu^*$)
which reproduce the experimental data $T_c=30$ K and $\alpha_{\rm C}=0.21$.

\begin{figure}
\centerline{\psfig{figure=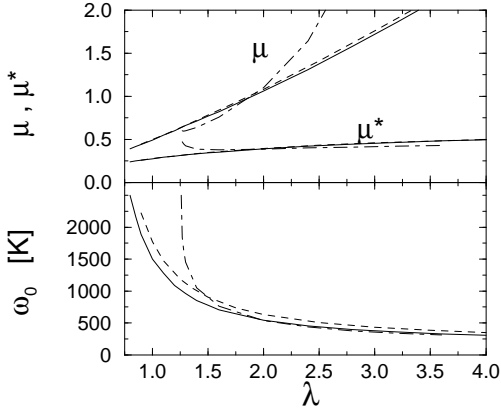,width=6.6cm}}
\caption{Plot of the Coulomb parameters $\mu$ and $\mu^*$ (upper panel) 
and of the central phonon frequency $\omega_0$ (lower panel) 
obtained by the standard ME equations constrained to have $T_c=30$ K
and $\alpha_{\rm C}=0.21$.
Solid lines: $\Delta\omega_0/\omega_0=0$, $\omega_c = 5 \omega_{\rm max}$;
dashed lines: $\Delta\omega_0/\omega_0=1$, $\omega_c = 5 \omega_{\rm max}$;
dot-dashed lines: $\Delta\omega_0/\omega_0=0$, $\omega_c = E_{\rm F}$.}
\label{fig1}
\end{figure}

In Fig. \ref{fig1} we show the calculated $\mu$ and $\mu^*$ (top panel)
and $\omega_0$ (lower panel) as function of $\lambda$ 
for $\Delta\omega_0/\omega_0=0$ (solid lines) and
$\Delta\omega_0/\omega_0=1$ (dashed lines) with 
$\omega_c=5\omega_{\rm max}$.
The main point of Fig. \ref{fig1} is that the calculated $\omega_0$ depends
strongly on the electron-phonon constant $\lambda$. 
For large values of $\lambda$, $T_c=30$ K and $\alpha_{\rm C}=0.21$
are reproduced only for quite small phonon frequencies while decreasing
$\lambda$ quickly enhances $\omega_0$.
The C$_{60}$ phonon spectrum however is limited by a maximum frequency
of $\sim 2300$ K \cite{gunnarsson,hebard}
and this settles a lower limit for the allowed values of $\lambda$.
In fact, from the lower panel of Fig. \ref{fig1},
it is clear that for $\lambda$ smaller than $1$ the corresponding 
$\omega_{\rm max}=\omega_0+\Delta\omega_0/2$ rapidly exceeds $2300$ K
signalling that the solution of the ME equations falls well outside the
range of applicability for the fullerene compounds.

A further interesting point is that
the effect of the spectral broadening ($\Delta\omega_0>0$) is
of secondary importance to the overall behavior, 
suggesting that the results are only weakly affected
by the detailed structure of $\alpha^2\! F(\omega)$. In fact, for $\lambda=0.9$
we find $\mu^*\simeq 0.25$, $\omega_{\rm ln}\simeq 1873$ K 
($\Delta\omega_0=0$) and $\omega_{\rm ln}\simeq 1555$ K 
($\Delta\omega_0=\omega_0$), where 
$\omega_{\rm ln}=\exp[(2/\lambda)\int d\omega \ln\omega
\alpha^2\! F(\omega)/\omega]$ is the logarithmic phonon frequency which,
for the rectangular model here considered,
is given by $\omega_{\rm ln}=[\omega_0^2-(\Delta\omega_0/2)^2]^{1/2}$.
These values are consistent with those reported in Ref. \cite{fuhrer}
($\lambda\simeq 0.9$, $\mu^*\simeq 0.22$, $\omega_{\rm ln}\simeq 1360$ K) 
where a phonon spectrum obtained by {\it ab initio} calculations
has been used to fit $T_c=30$ K and $\alpha_{\rm C}=0.21$.

In Fig. \ref{fig1} we show also the results for $\Delta\omega_0=0$
and $\omega_c=E_{\rm F}$ (dot-dashed lines). Due to the reduced dynamical
screening of the Coulomb repulsion, the increase of $\omega_0$ 
is much steeper than the previous cases 
until at $\lambda\simeq 1.25$ we find  $\omega_0\simeq E_{\rm F}$ and
$\mu^*\simeq \mu$ so that it is no longer possible to have
$\alpha_{\rm C}$ smaller than the BCS value $0.5$.

We now address the consistency of the data of Fig. \ref{fig1} with the
standard ME theory \cite{migdal,elia}. 
The assumption at the basis of the ME framework is
Migdal's theorem which states that, as long as the phonons have a much 
slower dynamics than that of the electrons, the nonadiabatic electron-phonon 
interference effects (vertex processes) can be neglected \cite{migdal}. 
This condition is well
satisfied in conventional superconductors since their Fermi energy is of order
$E_F\sim 10$ eV $\sim 10^5$ K while the highest phonon frequencies are
usually less than $\sim 50$ meV $\sim 60$ K \cite{carbotte}. However, 
the alkali-doped fullerene compounds are molecular solids characterized by
very narrow conduction electron bands of width of
only $W\simeq 0.5$ eV \cite{gunnarsson}.
In Rb$_3$C$_{60}$ (as in the other A$_3$C$_{60}$ compounds) the conduction
band is half-filled by electrons and the corresponding Fermi energy is
$E_F\simeq 0.25$ eV $=2900$ K, while the maximum phonon frequency
is $\sim 2300$ K. In principle, therefore, there is no reason
to expect Migdal's theorem to be applicable in fullerene compounds, unless
the main interaction is with the lowest C$_{60}$ phonon modes ($\sim 400$ K) 
via however a rather weak coupling.

\begin{figure}
\centerline{\psfig{figure=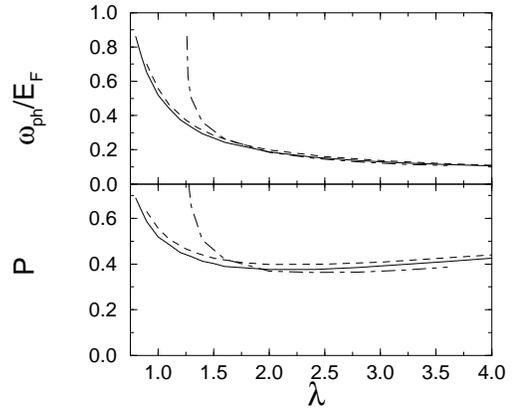,width=6.6cm}}
\caption{Values of $\omega_{\rm ph}/E_F$ (upper panel) and
of $P=\lambda\omega_{\rm ph}/E_F$ (lower panel) extracted from
the data of Fig. 1 with $E_F=0.25$ eV$=2900$ K. Solid lines: Einstein
phonon spectrum; dashed lines: broad spectrum with
$\Delta\omega_0/\omega_0=1$, dot-dashed lines: Einstein
phonon spectrum $\Delta\omega_0/\omega_0=1$ with $\omega_c = E_{\rm F}$.
The Migdal-Eliashberg theory holds
true only when $\omega_{\rm ph}/E_F\ll 1$ and $P\ll 1$.}
\label{fig2}
\end{figure}

We can test whether the data of Fig. \ref{fig1} are consistent with Migdal's
theorem by evaluating the order of magnitude ($P$) of the first nonadiabatic
electron-phonon vertex correction. By following the same reasonings 
of Migdal \cite{migdal,psg}, this is given by:
\begin{equation}
\label{migdal}
P=2\int_0^{+\infty}\! d\omega\frac{\alpha^2\! F(\omega)}{E_F}\equiv
\lambda\frac{\omega_{\rm ph}}{E_F},
\end{equation}
where for the rectangular model we considered the average phonon
frequency $\omega_{\rm ph}$ reduces to
\begin{equation}
\label{ave1}
\omega_{\rm ph}=\frac{2}{\lambda}\int_0^{+\infty}\! 
d\omega\,\alpha^2\! F(\omega)=
\frac{\Delta\omega_0}{\ln\left(\frac{\displaystyle \omega_0+\Delta\omega_0/2}
{\displaystyle \omega_0-\Delta\omega_0/2}\right)}.
\end{equation}
When $P\rightarrow 0$ the nonadiabatic interferences
are negligible and the ME theory holds true;
on the contrary,
sizeable values of $P$ signal the breakdown of the standard theory.
In Fig. \ref{fig2} we show the values of $\omega_{\rm ph}/E_F$ (up\-per panel)
and of $P$ (low\-er panel)  extracted from the $\omega_0$ values reported in 
Fig. \ref{fig1} and by setting $E_F=2900$ K.
As expected from the overall trend of $\omega_0$ {\it vs} $\lambda$ plotted
in Fig. \ref{fig1},
the adiabatic ratio $\omega_{\rm ph}/E_F$ is large and close to unity for $\lambda <1$,
while it rapidly decreases to $\omega_{\rm ph}/E_F\simeq 0.1$ for large values of
$\lambda$. One could therefore argue that adiabaticity is guaranteed
for very large electron-phonon couplings and that in this regime the ME framework is valid.
This is however incorrect because according to (\ref{migdal}) the vertex 
correction is proportional to $\lambda$ so that, as shown in the lower panel of 
Fig. \ref{fig2}, $P$ {\it is never negligible}. Note that the claimed value 
$\lambda\simeq 0.9$ \cite{fuhrer} corresponds to 
$\omega_{\rm ph}/E_F > 0.7$ 
and a minimum $P > 0.6$, {\it i.e.},
the vertex correction is comparable to unity.

From Figs. \ref{fig1} and \ref{fig2} we conclude therefore that the conventional
phonon-mediated superconductivity is not a complete and self-consistent picture
of Rb$_3$C$_{60}$ since the values of $\lambda$ and $\omega_0$ needed to
fit $T_c=30$ K and $\alpha_{\rm c}=0.21$ strongly violate Migdal's theorem.
This conclusion holds true even when electron-phonon spectra more structured
than the rectangular one are used to fit the data of Ref.\cite{fuhrer}.
By adding a $\delta$-peak centered at $\omega=\omega_1$ to a rectangular 
spectrum of width covering the whole intramolecular modes,
we have in fact simulated additional contributions from low-frequency
($\omega_1\simeq 50$ K) C$_{60}$-C$_{60}$ phonon modes \cite{mazin},
and from an enhanced coupling to soft intramolecular modes \cite{prassi}
($\omega_1\simeq 400-600$ K) possibly related to dynamical Jahn-Teller 
effects \cite{loktev}.
We find that $T_c=30$ K and $\alpha_{\rm C}=0.21$ imply $P > 0.4$ when
$\omega_1=50$ K and $P > 0.45$ when $\omega_1=400$ K
(further details will be presented elsewhere).

The above results point out that, if superconductivity in Rb$_3$C$_{60}$
is mediated by phonons, a consistent description of its superconducting properties
should be sought beyond the ME theory.
More precisely, the low value of $E_F$ indicates that the adiabatic hypothesis
and Migdal's theorem should be abandoned from the start and that the theory should
be formulated by allowing $\omega_{\rm ph}/E_F$ to have values sensibly
larger than zero. This naturally leads to
nonadiabatic interference effects in the electron-phonon scattering
which can significantly modify
both the normal and superconducting properties 
with respect to the ME phenomenology \cite{gpsprl}.

Indeed, characteristic effects of the nonadiabatic vertex corrections 
are predicted to be observable in several quantities, like 
$T_c$ and its the isotope coefficient \cite{gpsprl,psg}, 
the reduction rate of $T_c$ itself upon disorder \cite{sgp}, 
the effective electronic mass $m^*$ \cite{gcp}, 
the Pauli susceptibility \cite{gp}.
A peculiar feature of the nonadiabatic processes is 
to produce, under some conditions,
constructive electron-phonon interference in the particle-particle channel
leading to an enhancement of $T_c$ \cite{gpsprl,psg}.
Hence $T_c=30$ K, for
a given phonon spectrum, can be achieved by much smaller values
of $\lambda$ than needed in conventional ME theory.
Favourable conditions to this trend are expected in materials with
strong electronic correlation, as fullerenes:
strong local repulsion suppresses short-range interactions (large
${\bf q}$'s in Fourier space) and favours forward small-${\bf q}$
scattering \cite{zeyher,grilli} for which electron-phonon vertex 
processes become attractive.

Now we re-analyze the experimental constraints
of Rb$_3$C$_{60}$, $T_c=30$ K and
$\alpha_{\rm C}=0.21$, in the context of the nonadiabatic theory of
superconductivity. We show that the inconsistencies
of the results derived by ME theory are naturally solved when
the same experimental data are coherently analyzed in the 
nonadiabatic regime.

Explicit analytical and diagrammatic equations of the nonadiabatic
theory of superconductivity have been outlined in previous
works \cite{gpsprl,psg,sgp} and,
for sake of shortness, they will be here omitted.
A set of generalized Eliashberg equations in nonadiabatic regime
is constructed by following a perturbative approach based on
the Migdal parameter $P$. The consistency of such a perturbative
scheme is discussed below.
We simplify the phonon spectrum by assuming a dispersionless
Einstein phonon with energy $\omega_0$.
We note however that our results,
as also shown in Figs. \ref{fig1} and \ref{fig2},
are only weakly affected by the specific shape of $\alpha^2\! F(\omega)$.
Electronic correlation is taken into account by a cut-off $q_c$ on
the electron-phonon exchanged momenta, which selects forward scattering,
where stronger the correlation smaller $q_c$. 

\begin{figure}
\centerline{\psfig{figure=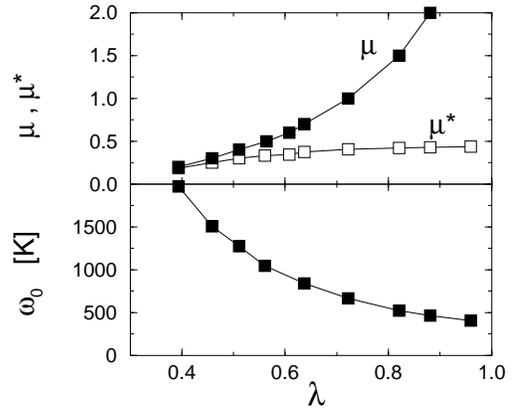,width=6.6cm}}
\caption{Plot of $\mu$, $\mu^*$ 
and $\omega_0$ 
calculated by the nonadiabatic theory as solutions of
$T_c=30$ K and $\alpha_{\rm C}=0.21$. Note that the range of values
for $\lambda$ is now much smaller and realistic with respect
to those of Fig. \ref{fig1}.}
\label{fig3}
\end{figure}

In Fig. \ref{fig3} we show the phonon frequency $\omega_0$, 
the statically  and dinamically screened Coulomb repulsion, 
respectively $\mu$ and $\mu^*$,
{\it vs.} $\lambda$ obtained in the nonadiabatic theory to
reproduce $T_c=30$ K and $\alpha_{\rm C}=0.21$. The parameter
$q_c$ has been chosen $q_c = 0.2 k_{\rm F}$ ($k_{\rm F}$ the Fermi vector),
an appropriate value for a strongly correlated system, and
the dinamical screening of $\mu$ has been considered to be provided
by the $t_{\rm 1u}$ electrons \cite{koch}.

From the comparison of Fig. \ref{fig3} with Fig. \ref{fig1},
a first remarkable difference lies in the distinct ranges
of electron-phonon couplings needed to reproduce the Rb$_3$C$_{60}$ data.
In fact, the conventional ME theory predicts $\lambda\bsim 1$ (Fig. \ref{fig1}), while
the nonadiabatic analysis yields $\lambda\lsim 1$ (Fig. \ref{fig3}).
But the most striking difference is that,
if we now take the parameters obtained by the generalized theory
and use them in the standard ME theory, these would give a very 
low value of $T_c$, less than $1$ K or even zero.
This result is now perfectly compatible with the GIC superconductors,
for which Migdal's theorem holds true,
and it clarifies that the high $T_c$ values
of the fullerides are essentially due to constructive nonadiabatic interference
effects rather than to a very large value of $\lambda$.
In our perspective, therefore, the origin of
the enhancement from $T_c\simeq 0.2$ K in GIC to $T_c\simeq 20-30$ K
in fullerene compounds stems mainly from the opening the nonadiabatic
channels in the electron-phonon interaction, rather than from a $\sim 300$\%
enhancement of $\lambda$.

\begin{figure}
\centerline{\psfig{figure=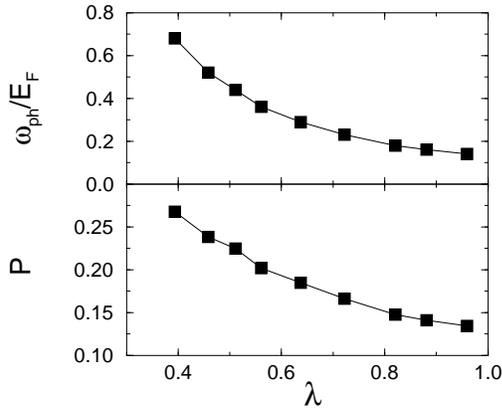,width=6.6cm}}
\caption{Adiabatic ratio $\omega_{\rm ph}/E_F$ and vertex correction
magnitude $P$ obtained by the nonadiabatic solutions of Fig. \ref{fig3}}
\label{fig4}
\end{figure}

We address now the consistency of the perturbative scheme with
the nonadiabatic solutions for Rb$_3$C$_{60}$.
In Fig. \ref{fig4} we show the adiabatic ratio
$\omega_{\rm ph}/E_F$ ($\omega_{\rm ph}=\omega_0$)
and the Migdal's parameter $P$ extracted from the data reported in Fig. \ref{fig3}.
The large value of $\omega_{\rm ph}/E_F$ shown in the upper panel
points out again the breakdown of Migdal's theorem and consequently
the need of the inclusion of the nonadiabatic vertex corrections.
The magnitude of the vertex corrections
$P \sim 0.2$, certainly not negligible, is however small enough to
support a perturbative approach in $P$ \cite{gpsprl,psg}.
Note moreover that, according to the comparison with exact results
for the single-electron Holstein model \cite{capone},  for weak couplings the system 
is away from polaron formation and that the perturbative scheme is well 
defined.

In conclusion, we have investigated the validity of Migdal-Eliashberg
theory of superconductivity in Rb$_3$C$_{60}$ by analyzing the
constraints imposed by recent experimental data, 
namely the critical temperature $T_c=30$ K
and the isotope effect $\alpha_{\rm C}=0.21$. 
We have found that the values of $\lambda$ and $\omega_{\rm ph}$
needed to reproduce the experimental data,
together with the very low value of the Fermi energy,
strongly violate Migdal's theorem and are therefore inconsistent with the
ME framework.
This situation unavoidably leads to the opening of
nonadiabatic channels in the electron-phonon pairing
which we argue to play the primary role for the high
values of $T_c$ in fullerene compounds. Finally, we stress the
importance of peculiar nonadiabatic effects in both 
superconducting \cite{sgp} and normal state properties \cite{gcp,gp} 
of fullerene compounds. Experiments in this direction are therefore
of great interest.

\end{document}